\RequirePackage{fix-cm}
\documentclass[twocolumn]{svjour3}          
\smartqed  
\usepackage{graphicx}
\usepackage{url}
\usepackage{amssymb}
\usepackage{amsmath}
\usepackage{graphicx}
\usepackage{epsfig}
\usepackage{graphics}
\usepackage[latin1]{inputenc}
\usepackage[numbers,sort]{natbib}
\usepackage[normalem]{ulem}

\journalname{General Relativity and Gravitation}

\RequirePackage{color}

\begin{document}

\title{Structure of neutron stars in $R$-squared gravity}

\author{Mariana Orellana $^{1,2,*}$  \and Federico Garc\'{\i}a $^{1,2,**}$ \and Florencia~A. Teppa Pannia $^{2,**}$ \and Gustavo~E. Romero $^{1,2,*}$}

\authorrunning{M. Orellana et al.}

\institute{ \at $^{*}$ Member of CONICET \at $^{**}$ Fellow of CONICET \at $^{1}$ Instituto Argentino de Radioastronom\'{\i}a CCT La Plata (CONICET), C.C.5, (1894) Villa Elisa, Buenos Aires, Argentina \at $^{2}$ Facultad de Ciencias Astron\'omicas y Geof\'{\i}sicas, Universidad Nacional de La Plata, Paseo del Bosque s/n, 1900 La Plata, Argentina
\\\email{morellana@fcaglp.unlp.edu.ar}
}

\date{Received: date / Accepted: 31/12/2012}

\maketitle

\begin{abstract}
The effects implied for the structure of compact objects by the modification of General Relativity produced by the generalization of the Lagrangian density to the form $f(R)=R+\alpha R^2$, where $R$ is the Ricci curvature scalar, have been recently explored. It seems likely that this {\it squared}-gravity may allow heavier Neutron Stars (NSs) than GR. In addition, these objects can be useful to constrain free parameters of modified-gravity theories. The differences between alternative gravity theories is enhanced in the strong gravitational regime. 
In this regime, because of the complexity of the field equations, perturbative methods become a good choice to treat the problem. Following previous works in the field, we performed a numerical integration of the structure equations that describe NSs in $f(R)$-gravity, recovering their mass-radius relations, but focusing on particular features that arise from this approach in the profiles of the NS interior. 

We show that these profiles run in correlation with the second-order derivative of the analytic approximation to the Equation of State (EoS), which leads to regions where the enclosed mass decreases with the radius in a counter-intuitive way. 
We reproduce all computations with a simple polytropic EoS to separate zeroth-order modified gravity effects.  

\end{abstract}

\section{Introduction}\label{intro}
Current cosmological observations interpreted in the standard cosmological model require the presence of a non-standard matter content in order to explain the accelerated expansion of the Universe \citep{Riess2009,Percival2010,Amanullah2010,Larson2011,Komatsu2011}. Along the last decade alternative cosmological models have been developed to reinterpret these data without involving any unknown, doubtful component of the energy-matter tensor (see, for instance, \cite{Bolejko2011}). The appearance of Extended Theories of Gravity (ETGs) was strongly stimulated by the possibilities they might provide in this context \citep{Sotiriou2010,deFelice2010,Capozziello2011}.
                             
ETGs are based on corrections and generalizations of Einstein's General Relativity (GR) theory. We focus on a particular class, called $f(R)$-gravity theories, which are based on a modification of the Einstein-Hilbert action: the usual Lagrangian density is generalized replacing the Ricci curvature scalar $R$ by a function of it or high-order invariants of the curvature tensor, such as $R^2$, $R_{\mu\nu}R^{\mu\nu}$, $R_{\mu\nu\alpha\beta}R^{\mu\nu\alpha\beta}$, $R\square R$, $R\square^kR$ (see \cite{Capozziello2011} for a complete review).  
 
Several of these models are successfully constructed to satisfy the current Solar System and laboratory tests \citep{Starobinski2007,HuSawicki2007,Miranda2009,Jaime2011}. In particular, the simplest choice $f(R)=R+\alpha R^{2}$, also called ``$R$-squared'' gravity, has been further studied as the basis for a viable alternative cosmological model, that can lead to the accelerated expansion of the Universe and is well consistent with the temperature anisotropies observed in Cosmological Microwave Background \citep{deFelice2010}.  
But in contrast to gravity in the weak-field regime, which has been subject to numerous experimental tests, gravity in the strong-field regime is largely unconstrained by observations \citep[e.g.][]{Dedeo2003}.

However, other authors, making a more detailed model of the structure of compact stars, obtained a set of modified Tolman-Oppenheimer-Volkoff (TOV) equations that describe a spherically symmetric mass distribution, under hydrostatic equilibrium, in simple $f(R)$-gravity, and derived solutions that reproduce the correct behaviour at the weak gravity limit \citep[e.g.][in the context of scalar-tensor theories of gravity]{Babichev2010}. 
Furthermore, working with a perturbative approach to solve the field equations, \citet{Cooney2010} and \citet{Santos2011} have found mass-radius relations for compact stars using a polytropic Equation of State (EoS). More recently \citet{Arapoglu2011} and \citet{Deliduman2011} applied the same approach using a set of realistic EoS for Neutron Stars (NSs) in the $R$-squared and $R_{\mu\nu}R^{\mu\nu}$ gravities, respectively. 

The mass-radius relations obtained by \citet{Arapoglu2011} and \citet{Deliduman2011} indicate that such $f(R)$ models can accommodate NSs up to masses larger than the currently observed ones, which are at most $M_{\rm max}= 1.97\pm 0.04 M_\odot$ for PSR J1614-2230 \citep{Ozel2010}. The $R$-squared gravity introduces a new parameter in the model through the value of $\alpha$, the coefficient of $R^2$. The freedom in the choice of this parameter allows some EoS, which are excluded within the framework of GR, to be reconciled with the observations.
Motivated by these results, we investigate in detail the structure of NSs under this model for two different EoS. One of them is a polytropic approximation that we use here to separate zeroth-order modified-gravitational effects, whereas the other provides a realistic representation of nuclear matter at high densities.

The paper is organized as follows. In Section 2, we obtain the modified TOV equations following the perturbative approach to solve the field equations. In Section 3, we present the EoS used to integrate the equations and we briefly describe the numerical methods of resolution. In Section 4, we present our results for the mass-radius relations, focusing on the behaviour of the profiles obtained for the NS interior. Final remarks are shown in Section 5.


\section{Structure equations in $R$-squared gravity} \label{model}

The equations that define the structure of a NS in GR are deduced proposing the static, spherically symmetric line element, d$s$, to be
\begin{equation}
{\rm d}s^2=-e^{2 \Phi(r)} c^2 {\rm d}t^2 + e^{2 \Lambda(r)} {\rm d}r^2+ r^2({\rm d}\theta^2+\sin^2{\theta} {\rm d}\varphi^2).
\label{metrica}
\end{equation}
Then, considering the Einstein equations for an ideal energy-momentum tensor $T^{\mu}_{\nu}=\textrm{diag}\{-\rho c^2,p,p,p\}$, these equations, so-called TOV because of the pioneer work of \cite{Tolman1939} and \cite{Oppenheimer1939}, can be expressed as:
\begin{eqnarray}
\label{eq:masa}
\frac{{\rm d}m(r)}{{\rm d}r} & = & 4\pi r^2 \rho(r) \label{mtov-GR}\\
\frac{{\rm d}p(r)}{{\rm d}r} & = & \frac{c^2\rho(r)+p(r)}{2Gm(r)-c^2 r}\, G\left(\frac{4 \pi}{c^2}r^2 p(r) + \frac{m(r)}{r}\right)\\
e^{-2\Lambda(r)} & = & 1 -\frac{2Gm(r)}{c^2 r}\\
\frac{{\rm d}\Phi(r)}{{\rm d}r} & = &-\frac{1}{(c^2 \rho(r) + p(r))}\frac{{\rm d}p}{{\rm d}r} , \label{Phitov-GR}
\end{eqnarray}
\noindent where  $m(r)$ is the total relativistic mass enclosed in a sphere of radius $r$. The functions $\rho(r)$ and $p(r)$ are, respectively, the mass-energy density and the pressure at this radius. We explicitly keep the gravitational constant, $G$, and the velocity of light, $c$, since the quantities are considered with their full-dimensions for integration purposes.
Giving an explicit relation between $\rho$ and $p$, the so-called EoS, the TOV equations can be solved assuming a value for the central density, $\rho(r=0)=\rho_c$, and integrating the system to the pressure vanishes, $p(r=R_{\star})=0$. Here   $R_{\star}$ is the radius of the star and $M_{\star}=m(R_{\star})$  the stellar mass. Every $\rho_c$ generates a couple of values $M_{\star}$ and $R_{\star}$. Then, varying this parameter, a mass-radius $(M_{\star}-R_{\star})$ relation is defined for every EoS.

The modified TOV equations can be obtained from the gravitational field equations. Adding the new term to the Hilbert-Einstein action, we have:  
\begin{equation}
S=\frac{c^4}{16\pi G}\int {\rm d}^4x\sqrt{-g}(R + \alpha R^2)+S_{\rm matter},
\label{action}
\end{equation}
where $g$ is the determinant of the metric $g_{\mu\nu}$. 

Working in the metric formalism, the variation of the action with respect to the metric yields fourth-order differential equations of $g_{\mu\nu}$. This poses an enormous obstacle to solve the problem thoroughly.  
For this reason, we adopt the perturbative approach presented by \citet{Cooney2010} and \citet{Arapoglu2011}. Rewriting $f(R)=R + \alpha R^2=R(1+\beta)$, we consider the $f(R)$ function as a perturbation of a GR background. Hence, the dimensionless quantity $\beta\equiv\alpha R$ comprises the deviation from GR and the perturbative method can be applied as long as $|\beta| \ll 1$. 
Under this condition, we can work with equations of motion up to first order in $\beta$ without imposing any constrain at the level of the action and ensuring the nature of the variational principle \citep{Cooney2009}. 
Neglecting terms with $\mathcal{O} (\beta^2)$ or higher,
the field equations reduce, for this particular choice of $f(R)$, to
\begin{equation}\begin{split}
G_{\mu\nu} + \alpha &\bigg[2R \left(R_{\mu\nu}-\frac{1}{4}Rg_{\mu\nu}\right) +\cr
&+2\left(g_{\mu\nu}\square R - \nabla_{\mu}\nabla_{\nu}R\right)\bigg]=\frac{8\pi G}{c^4} T_{\mu\nu},
\end{split}
\label{field-eq}
\end{equation}
where $G_{\mu\nu}=R_{\mu\nu}-\frac{1}{2}Rg_{\mu\nu}$ is the Einstein tensor. In the limit $\alpha \longrightarrow 0$,  $\beta \longrightarrow 0$, and the field equations of GR are recovered \citep[see e.g.][]{Capozziello2011}.

We also assume a static and spherically symmetric line element, given by equation (\ref{metrica}). The perturbative approach allows to expand the functions present in the metric into a leading term (unperturbed), denoted with subscript 0, plus a corrective one, denoted with subscript 1, that is of first order in $\beta$: $\Lambda=\Lambda_0+\beta \Lambda_1$ and $\Phi=\Phi_0+\beta \Phi_1$. The hydrodynamic quantities are also defined perturbatively:  $\rho=\rho_0+\beta\rho_1$ and $p=p_0+\beta p_1$ \citep{Arapoglu2011}. 
Hence, new restrictions are imposed over the value of $\beta$ by the constraints $\beta \Phi_1\ll\Phi_0$, $\beta \Lambda_1\ll\Lambda_0$, $\beta \rho_1\ll\rho_0$, and $\beta p_1\ll p_0$. 
From now on, we use the prime for radial derivatives. 

Following \cite{Cooney2009} and \cite{Arapoglu2011}, we define the mass assuming that the solution for the metric has the same form as the exterior Schwarzschlid solution in GR, i.e.
\begin{equation}
e^{-2\Lambda(r)}=1-\frac{2GM_{\ast}}{c^2r}, \mbox{  for } r>R_\star.
\end{equation}
For the interior solution,
\begin{equation}
\label{2.11}
e^{-2\Lambda(r)}=1-\frac{2Gm(r)}{c^2r}, \mbox{  for } r<R_\star,
\end{equation}
where $m$ also admits a perturbative expansion $m=m_0+\beta m_1$, with $m_0$ the zeroth-order mass that is obtained integrating (\ref{mtov-GR}).

With this considerations, and taking into account that $\rho_0$ and $p_0$ satisfy Einstein's equations, the derived modified TOV equations 
are:
\begin{equation}
\begin{split}
&\!\!\frac{{\rm d}m}{{\rm d}r}= 4 \pi r^2 \rho-2 \beta \bigg[\overbrace{4 \pi r^2 \rho_0 -\frac{c^2}{8 G}r^2R_0}^{A/2} +\cr
&\!\!\!+\!\underbrace{\left(2 \pi \rho_0 r^3 - \frac{c^2}{G}r + \frac{3}{2}m_0\right)\frac{R_0'}{R_0}}_{B/2} -\underbrace{r\left(\frac{c^2}{2G}r-m_0\right)\frac{R_0''}{R_0}}_{-C/2}\bigg]\!\!.
\end{split}
\label{Mtov}
\end{equation}

\begin{equation}
\label{Ptov}
\begin{split}
\!\frac{{\rm d}p}{{\rm d}r}&=\frac{c^2\rho+p}{2Gm-c^2 r}\, G\bigg\{\left(\frac{4 \pi}{c^2}r^2 p + \frac{m}{r}\right)-
2 \beta \bigg[\frac{4\pi}{c^2}r^2 p_0 +\cr
&+\frac{c^2}{8G}r^2 R_0 + \left(\frac{2\pi}{c^2}p_0 r^3  + \frac{c^2}{G}r -\frac{3}{2}m_0 \right)\frac{R_0'}{R_0} \bigg]\bigg\}.
\end{split}
\end{equation}
Note that $2\beta\Big[... \Big]$ indicate the first order correction in $\beta$ into the gradients ${\rm d}m/{\rm d}r$ and ${\rm d}p/{\rm d}r$, respectively. In order to work up to first order in $\beta$, terms between the brackets have been evaluated at order zero. It is important to note that ${\rm d}p/{\rm d}r$ does not depend on $R_0''$. Mass-derivative terms are indicated, giving: ${\rm d}m/{\rm d}r= 4 \pi r^2 \rho-\beta \big[ A+B+C \big]$. We study each term contribution below.

Exact equations (\ref{Mtov}) and (\ref{Ptov}) are impractical because $\beta$ involves other derivatives through $R$, which makes $\beta$ $r$-dependent in a complicated way\footnote{The Ricci scalar in terms of the functions of the metric (\ref{metrica}) is:
%
$$\!\!\!R\!=\!\frac{2\,e^{-2\Lambda}\!\!
 \left[r^2 \left\{\Phi' \Lambda'+ (\Phi')^2\!\!+ \Phi''\!\right\}\!+\!2r\left(\Phi' - \Lambda'\!\right) -e^{2\Lambda}\!\!+1\right]}{r^2}\!\!\!$$
}.
Thus, we integrate equations (\ref{Mtov}) and (\ref{Ptov}) assuming that $\beta$ is well approximated by $\beta\simeq\hat{\beta}\equiv\alpha R_0$, where $R_0$ is the Ricci scalar {\it locally} calculated in GR and $\alpha$ is a constant parameter with squared distance units, compatible with other authors approach \citep[for instance][]{Arapoglu2011}. Note that 
\begin{equation}
R_0= \frac{8 G \pi}{c^4} (\rho_0 c^2 -3 p_0),
\end{equation}
\noindent and then, contrary to the GR case, the derivatives of the EoS, ${\rm d}p/{\rm d}\rho$ and ${\rm d}^2p/{\rm d}\rho^2$, also enter into (\ref{Mtov}) and (\ref{Ptov}), through $R_0'$ and $R_0''$.

In $f(R)$-gravity the weight of the perturbation is adjusted by the value of the $\alpha$ parameter. In our work, we restrict ourselves to the constraints reported by \citet{Santos2010}, and references therein, which points to  $10^8$cm$^2<\alpha/3<10^{10}$cm$^2$, based on astronomical observations and nuclear experiments in terrestrial laboratories. 

\begin{figure*}[ht!]
\begin{center}
\includegraphics[width=8cm]{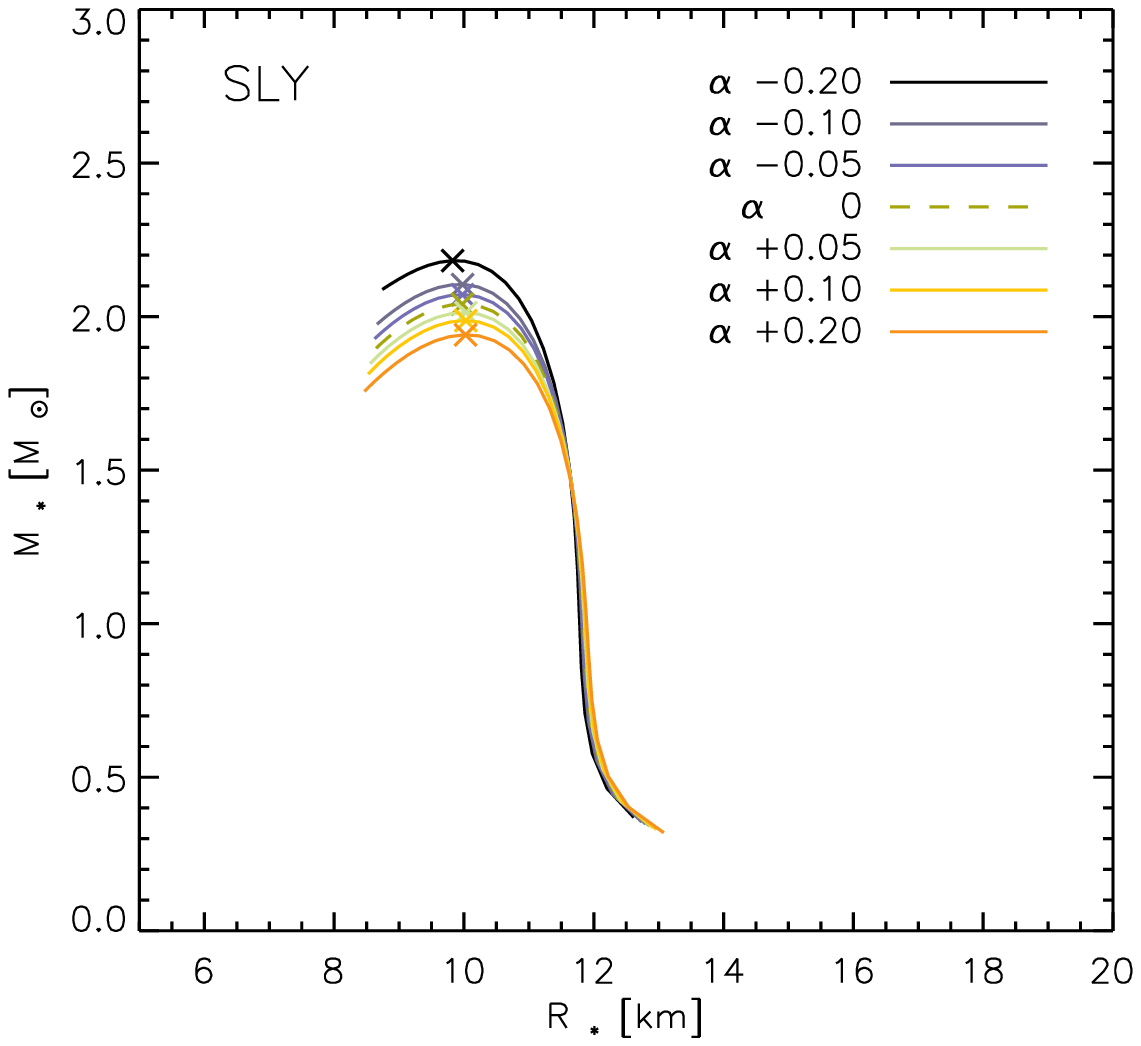}\hspace{1cm}\includegraphics[width=8cm]{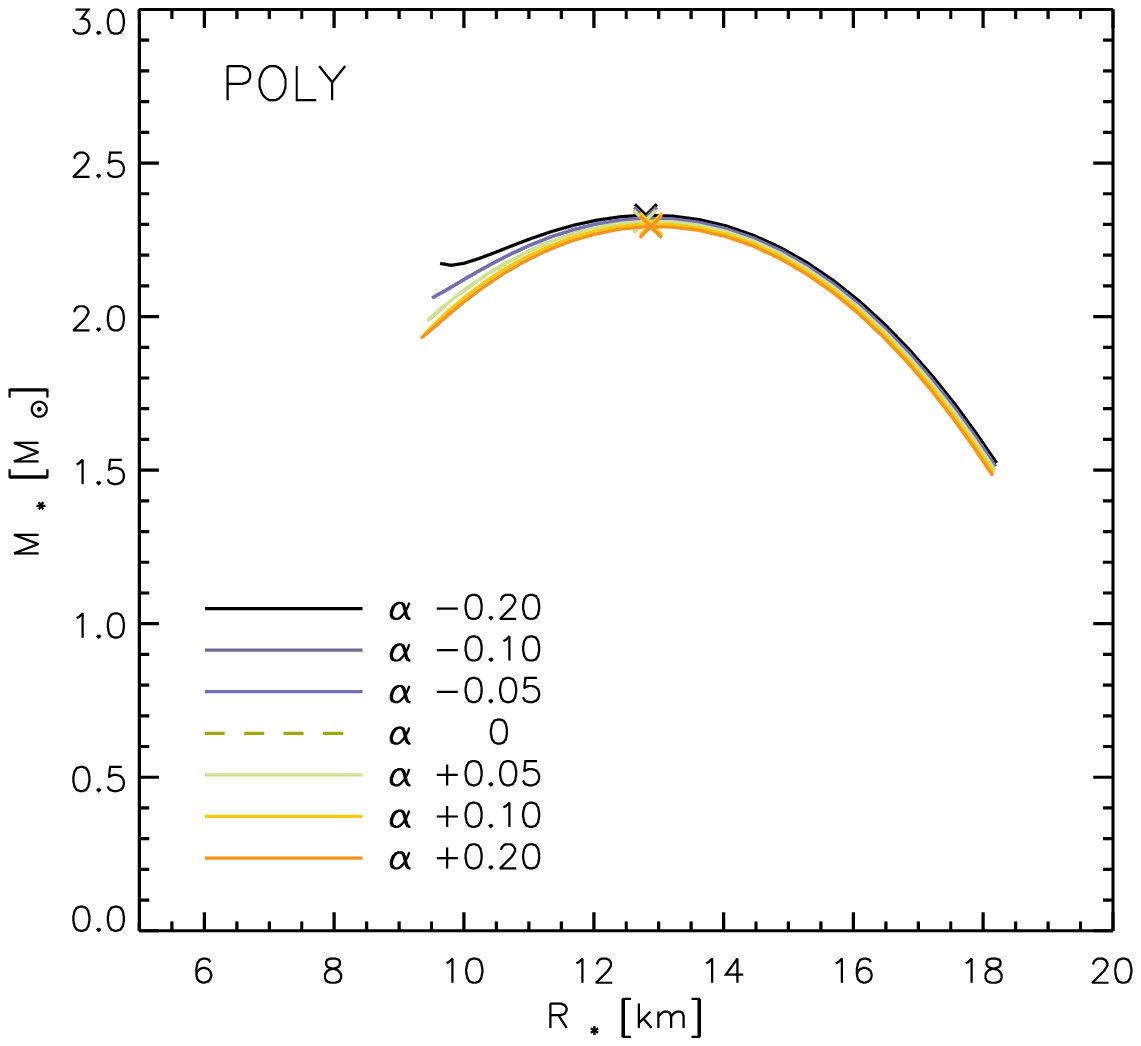}
\caption{Mass-radius $(M_{\star}-R_{\star})$ relations for the two selected EoS: SLY and POLY (left and right, respectively), considering seven values for the $\alpha$ parameter, which are indicated above in km$^2$ units. All the curves correspond to values of central density, $\rho_c$, in the range $10^{14.6}-10^{15.9}$~gr~cm$^{-3}$. The crosses indicate the maximum mass for each curve, assuming a necessary condition for equilibrium: ${\rm d}M/{\rm d}\rho_c > 0$.}
\label{configs-4}
\end{center}
\end{figure*}
\begin{figure*}
\begin{center}
\includegraphics[width=8cm]{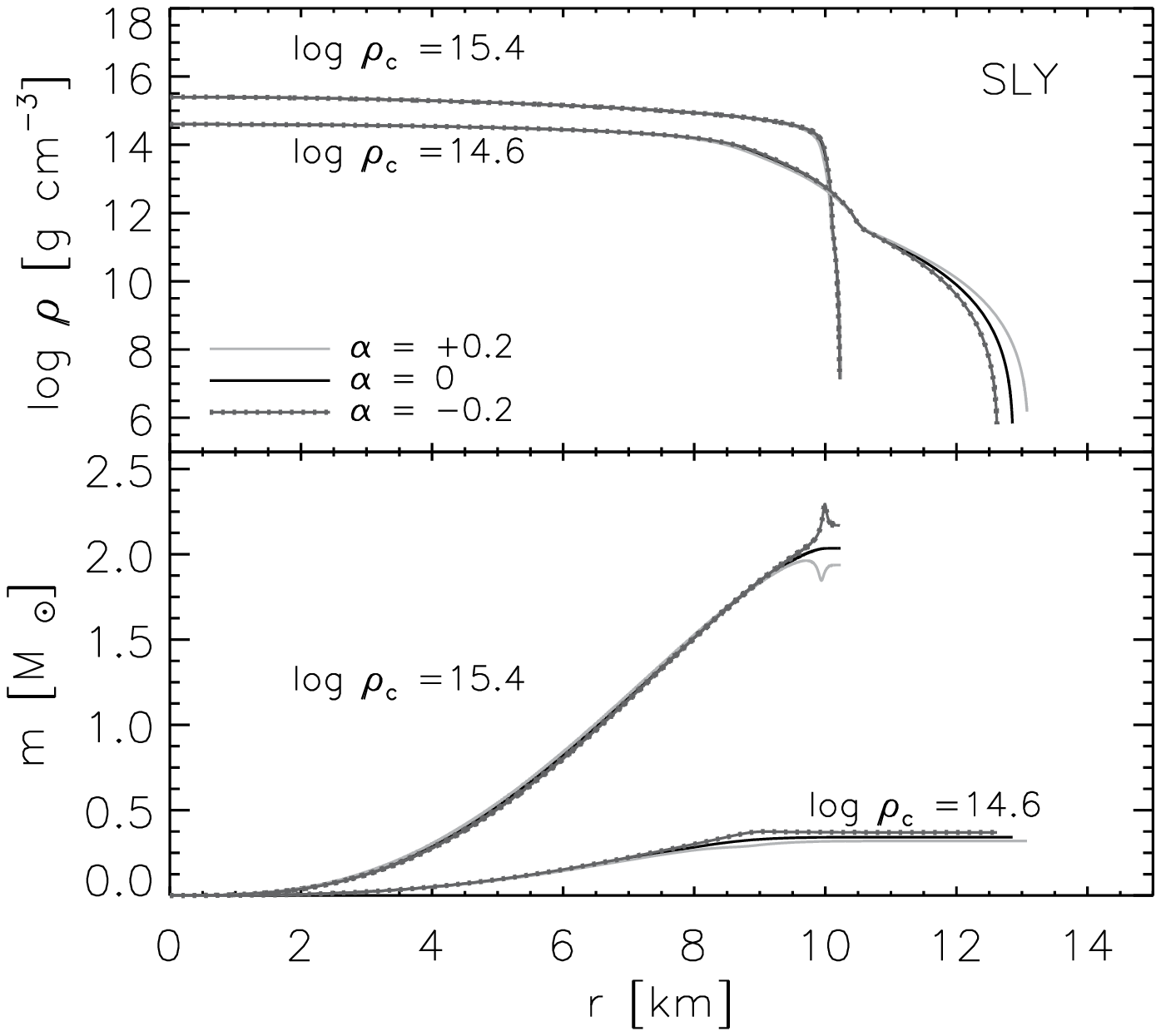}\hspace{1cm}\includegraphics[width=8cm]{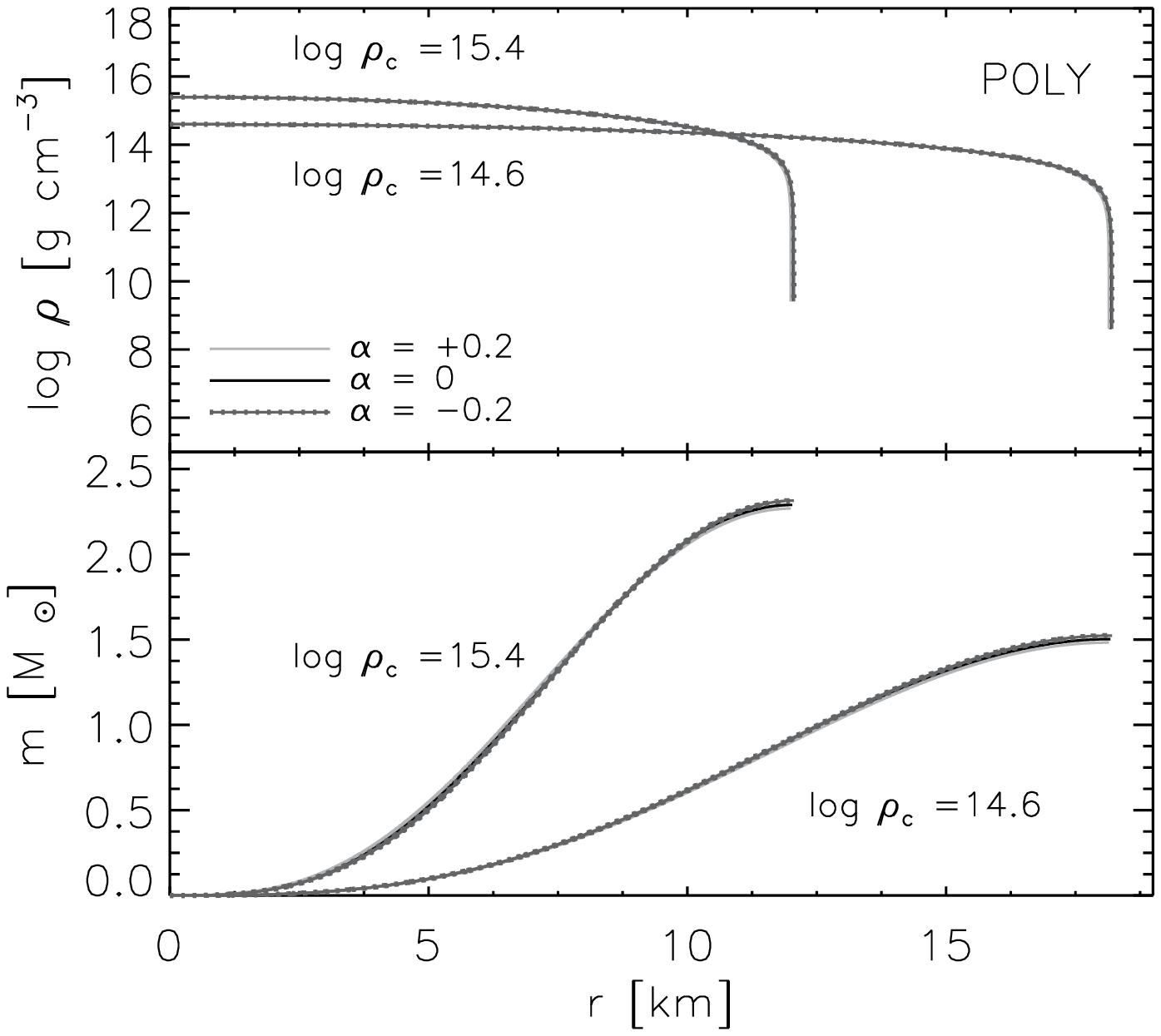}
\caption{Profiles of the internal structure of NSs for two extreme cases of low and high central densities, $\rho_c=10^{14.6}$ and $10^{15.4}$~gr~cm$^{-3}$, and for three different values of the $\alpha$ parameter (+0.2, 0.0 and --0.2 km$^2$), where $\alpha=0.0$ corresponds to GR case. On the left (right) panel profiles corresponding to the SLY (POLY) EoS are shown. A zoom-in of the mass profile close to the NS surface is shown in Figure \ref{perfil1_zoom}. At low central density values the effect on the integrated mass can still represent a deviation as much as $\le 10$\% from the GR mass.}
\label{perfil1}
\end{center}
\begin{center}
\resizebox{\hsize}{!}{\includegraphics{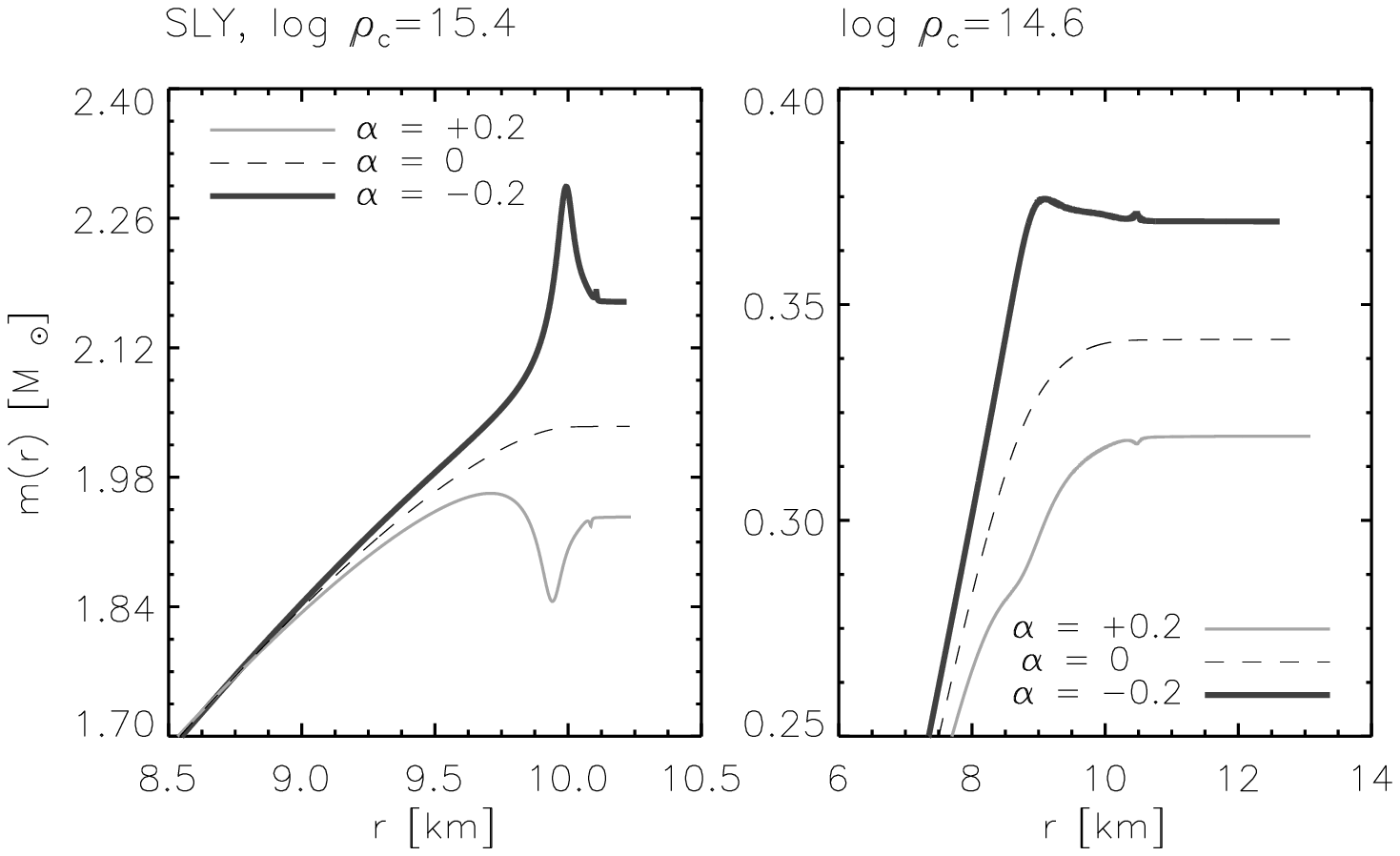} \includegraphics{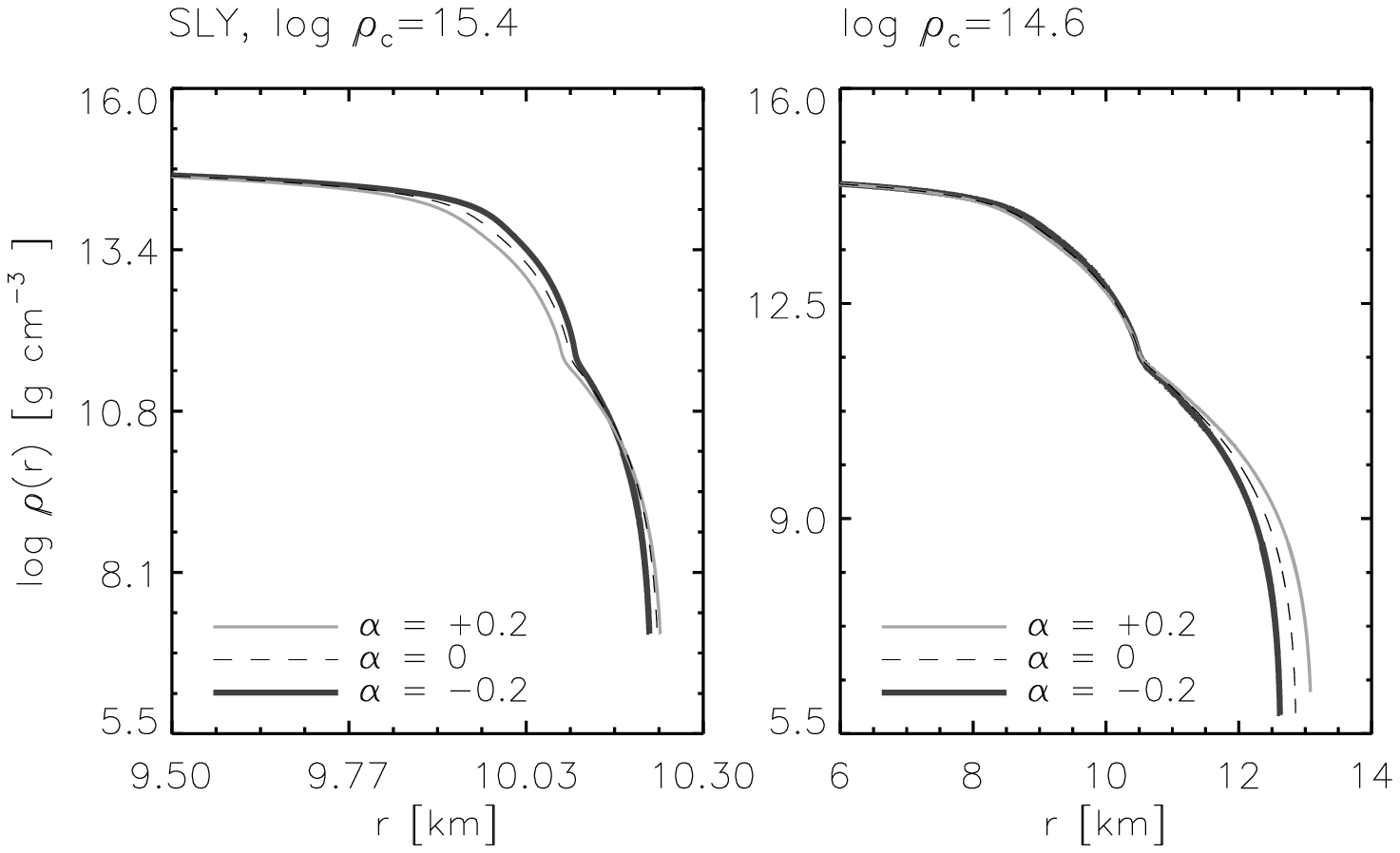}}
\caption{Zoom-in of the the mass and density profiles close of the surface of the NS for the SLY EoS shown in Figure \ref{perfil1}, for $\rho_c=10^{15.4}$ and $10^{14.6}$~gr~cm$^{-3}$. For the value of the $\alpha$ parameter --0.2 km$^2$ (+0.2 km$^2$) the mass increases (decreases) roughly $10\%$ respect to de GR case ($\alpha = 0$), for both low and high central densities.}
\label{perfil1_zoom}
\end{center}
\end{figure*}

\section{Equations of state and numerical methods}

To solve the system of equations given by (\ref{Mtov}) and (\ref{Ptov}) is necessary to bring a relation between the pressure $p$ and the density $\rho$ or the energy density $\varepsilon$, the so-called EoS. The EoS contains the information of the behaviour of matter inside NSs through several orders of magnitude in density. Because the properties of the matter at the highest densities in the central region of NSs are not well understood, different EoS are proposed and then constrained with observations of masses and radii of actual NSs.

An analytical representation of the EoS is required for solving the structure of NSs in modified theories of gravity, where hydrostatic equilibrium equations are of fourth-order. In such cases the usual interpolation technique fails to accurately represent high-order derivatives \citep{Gungor2011}.
Analytical representations of several EoS have been obtained through a consistent procedure by \citet{Haensel2004}, who calculated the best-fit coefficients of a polynomial expansion both in the crust and core density regimes.
However, it must be emphasized that these analytical EoS are approximations obtained by fitting only the zeroth-order relation between $\rho$ and $p$, because it is the relation needed to calculate the structure of NSs in GR. Thus, special care should be taken if high-order derivatives of these expressions are used during the calculation, as in the case we are interested in here (i.e. ${\rm d}p/{\rm d}\rho$ and ${\rm d}^2p/{\rm d}\rho^2$). 

Taking this into account, and in order to compare our results with those already published in the literature, we calculate mass-radius $(M_{\star}-R_{\star})$ relations considering two different EoS: SLY \citep{Douchin2001,Haensel2004} and POLY \citep{undergrad2004}. 

The first one is a realistic EoS that properly represents the behaviour of nuclear matter at high density. Its analytic parametrization is given by
\begin{eqnarray}
  \zeta&=&\frac{a_1+a_2\xi+a_3\xi^3}{1+a_4\,\xi}\,f_0(a_5(\xi-a_6))
\nonumber\\&&
     + (a_7+a_8\xi)\,f_0(a_9(a_{10}-\xi))
\nonumber\\&&
     + (a_{11}+a_{12}\xi)\,f_0(a_{13}(a_{14}-\xi))
\nonumber\\&&
     + (a_{15}+a_{16}\xi)\,f_0(a_{17}(a_{18}-\xi)) ~,
\label{eq:fit.P}
\end{eqnarray} 
where 
\begin{equation}
\xi=\log(\rho/\textrm{g cm}^{-3}),\,\,\,
\zeta = \log(P/\textrm{dyn}\,\textrm{cm}^{-2}),
\end{equation}
\begin{equation}
f_0(x) = \frac{1}{\mathrm{e}^x+1},
\end{equation}
and the coefficients $a_i$ are tabulated \citep{Haensel2004}. This is the same expression used by \cite{Arapoglu2011}, and we use it here to test our results.
The second EoS is a simpler polytropic approximation given by
\begin{equation}
\zeta=2\xi+5.29355 ~.
\end{equation}
Despite the latter is not a realistic EoS that thoroughly represents NSs, it is a {\it toy model} that allows to study zeroth-order modified gravity effects, separating them from more tricky effects arising in the case of a realistic EoS, with its complex analytical expression and from which the error propagating to the derivatives is out of our control.
The precise value of the adiabatic index, $\Gamma={\rm d}\log p/{\rm d}\log \rho={\rm d}\zeta/{\rm d}\xi$, is not relevant as $\Gamma$ remains a derivable function. The reader is referred to \citep{Read2009} for tighter constraints on $\Gamma$ that point to a somewhat larger value than the one adopted here.

\subsection{Numerical Method}

Solving the system of ordinary differential equations formed by the equations (\ref{Mtov}) and (\ref{Ptov}) implies their integration from the centre to the NS surface, using the chosen EoS. Once the solution is found, a couple of values for the mass, $M_{\star}$, and the radius, $R_{\star}$, are established. In order to perform the integration, we use a numerical code based on a fourth-order Runge-Kutta method on the radial coordinate. For this coordinate we implement a variable step which is systematically shortened close to the NS surface, to account for rapid variations of the physical parameters in this region.

During the Runge-Kutta loop, we also solve the differential equations corresponding to the metric components: $g_{tt}$ and $g_{rr}$, which are  
involved in the criteria for the validity of the perturbative approach. In each  step, we first integrate the TOV equations in the frame of GR to obtain zeroth-order values that then we use to calculate the first-order perturbative solution.

We start the numerical integration from the centre with a given central density, $\rho_c$, and we finish the integration at the surface, defining the NS radius, $R_{\star}$, and mass, $M_{\star}$, when the density reaches $\rho = 10^6$~gr~cm$^{-3}$. This density corresponds to the outer boundary of the NS crust and is the limit of validity for these kind of EoS, as they were conceived beginning with a model for nuclear matter at high densities.

\section{Results}
In Figure \ref{configs-4} we present the mass-radius $(M_{\star}-R_{\star})$ relations obtained for SLY and POLY EoS (left and right panels, respectively), using seven values of the $\alpha$ parameter between --0.2 and +0.2~km$^2$ and considering central densities, $\rho_c$, ranging from $10^{14.6}$~gr~cm$^{-3}$ to $10^{15.9}$~gr~cm$^{-3}$. Maximum masses achieved are indicated by crosses in each curve, assuming a necessary condition for equilibrium: ${\rm d}M/{\rm d\rho_c}>0$. Our results for SLY EoS are in accordance with those previously presented by \cite{Arapoglu2011}. Values of $\alpha < 0$~km$^2$ ($\alpha > 0$~km$^2$) can accommodate higher (lower) maximum masses than GR. In particular, POLY configurations are less sensitive to the value of $\alpha$ than those from SLY EoS.

In Figure \ref{perfil1} we present the internal profiles found for the density, $\rho(r)$, and mass, $m(r)$, for
 our $\alpha$ extreme values and the GR case.
The density and the pressure follow rather usual (resembling GR) profiles, where both magnitudes monotonously decrease with radius. 
However, particular effects of $f(R)$ are reflected in the mass profiles for the realistic SLY EoS, and become more pronounced for the high-mass NSs ($\rho_c=10^{15.4}$~gr~cm$^{-3}$).

These effects are evident close to the NS surface (at $r\sim10$~km) where, in a narrow layer ($\Delta r \sim 0.2$~km), an unexpected (counter-intuitive) decrease in $m(r)$ appears before ($\alpha>0$~km$^2$) or after ($\alpha<0$~km$^2$) a dip (peak) in the profile. In Figure \ref{perfil1_zoom} we present a zoom of the mass and density profiles (left and right panels, respectively) close to the NS surface, obtained using the SLY EoS for $\log \rho_c$~[gr~cm$^{-3}$] = 15.4 and 14.6. Besides the modification of $M_*$ is higher for high central density stars, the relative change in the total mass with respect to the GR case is roughly 10\% in both high/low central density cases.

In the frame of GR, a decreasing mass profile could only be accomplished by a fluid of negative density, because ${\rm d}m/{\rm d}r = 4 \pi r^2 \rho(r)$. However, in $f(R)$-gravity these profiles can be explained as a consequence of the modified geometry. In contrast, no such features are present in the profiles obtained with the POLY EoS for all the values considered for the parameters. 

\begin{figure}
\begin{center}
\resizebox{\hsize}{!}{\includegraphics{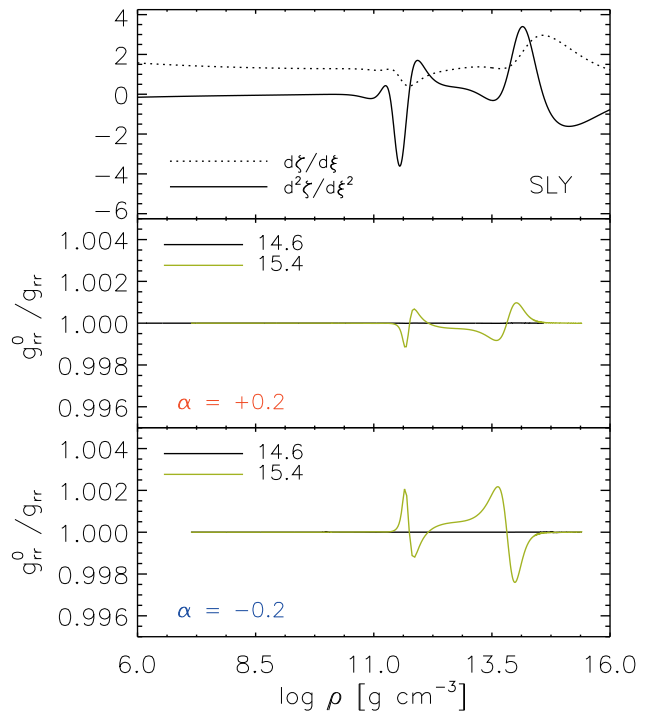}}
\caption{Top panel: Profile of the of the first (dotted line) and second (continuous line) logarithmic derivatives of the SLY EoS in the NS interior. Middle and Bottom panels: Profiles of the ratio between the radial component of the metric at order zero (GR), and at first order ($g_{rr}^0/g_{rr}$) for $\alpha=+0.2$ and $-0.2$~km$^2$, which should be close to $1.0$ as a necessary condition of the pertubative method. The perturbative deviations are closely related with the behaviour of the second-order derivative of the EoS.}
\label{perfil2}
\end{center}
\end{figure}

In Figure \ref{perfil2} we present the profile of the ratio $g_{rr}^0/g_{rr}$ with the density through all the NS interior for the SLY EoS, considering the same densities of Figure \ref{perfil1}. We recall that this ratio should be close to $1.0$ as a necessary condition for the validity of the pertubative method. On the upper panel of this figure we also plot the first and second logarithmic derivatives of the SLY EoS. From the comparison of the trend, a strong coupling of the perturbative deviations of $g_{rr}$ with the second-order derivative of the EoS is evident. 
The modification in the metric radial component is mild, and only perceptible when the second-order derivative becomes important, strongly oscillating, in the $10^{11}-10^{14}$~gr~cm$^{-3}$ density range. Such behaviour is not present for the POLY EoS, which logarithmic second derivative is null, maintaining $g_{rr}^0/g_{rr}\simeq 1$ all through the NS interior.
It is important to note that the function $\Phi(r)$ in the time component of the metric is actually reflecting the behaviour of the pressure, whereas $\Lambda(r)$ is governed by the mass, and it is the mass but not the pressure the one requiring evaluations of $R''_0$, and thus depending on high-order derivatives of the EoS.

\begin{figure*}
\begin{center}
\includegraphics[width=8cm]{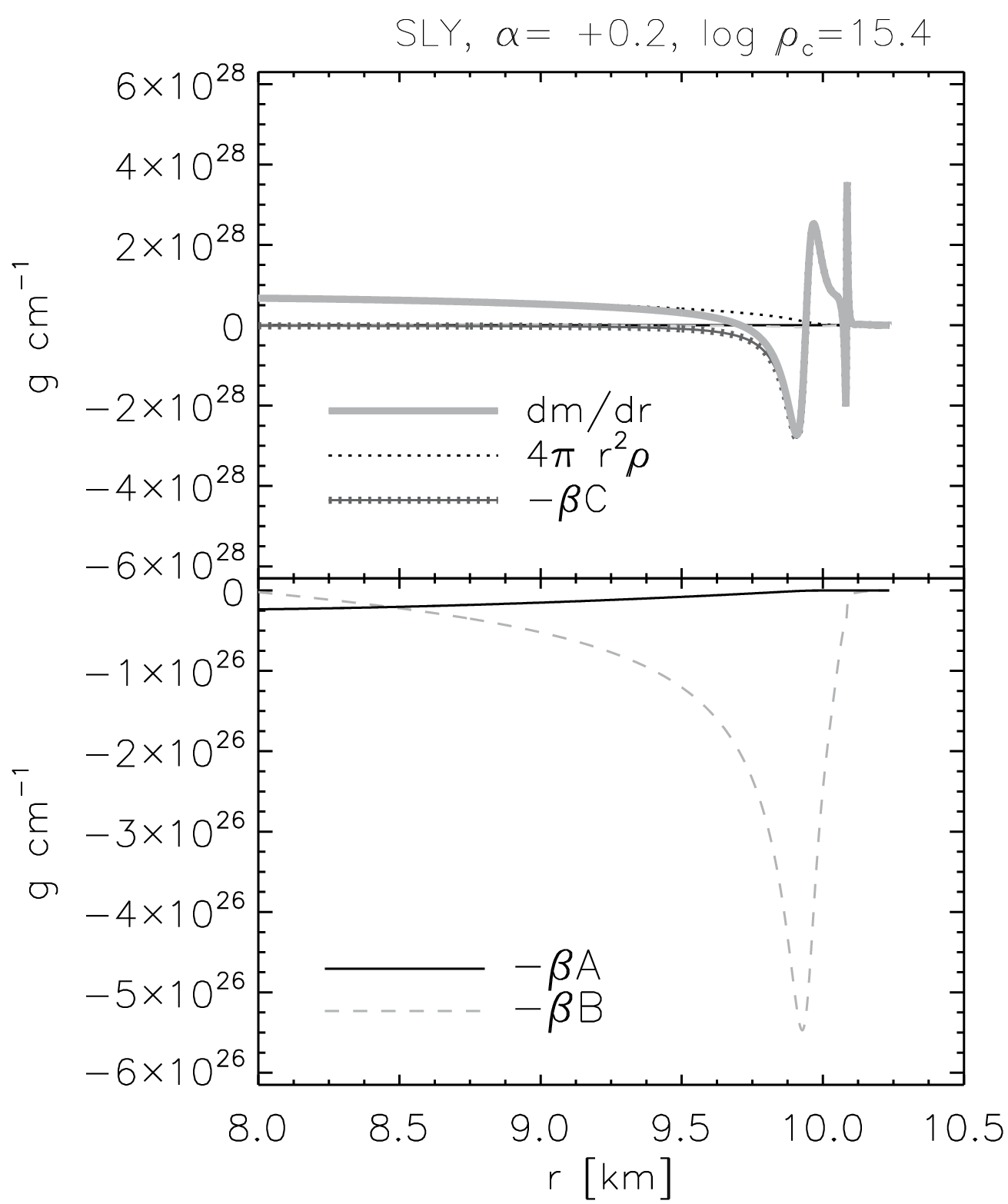}\hspace{1cm}\includegraphics[width=8cm]{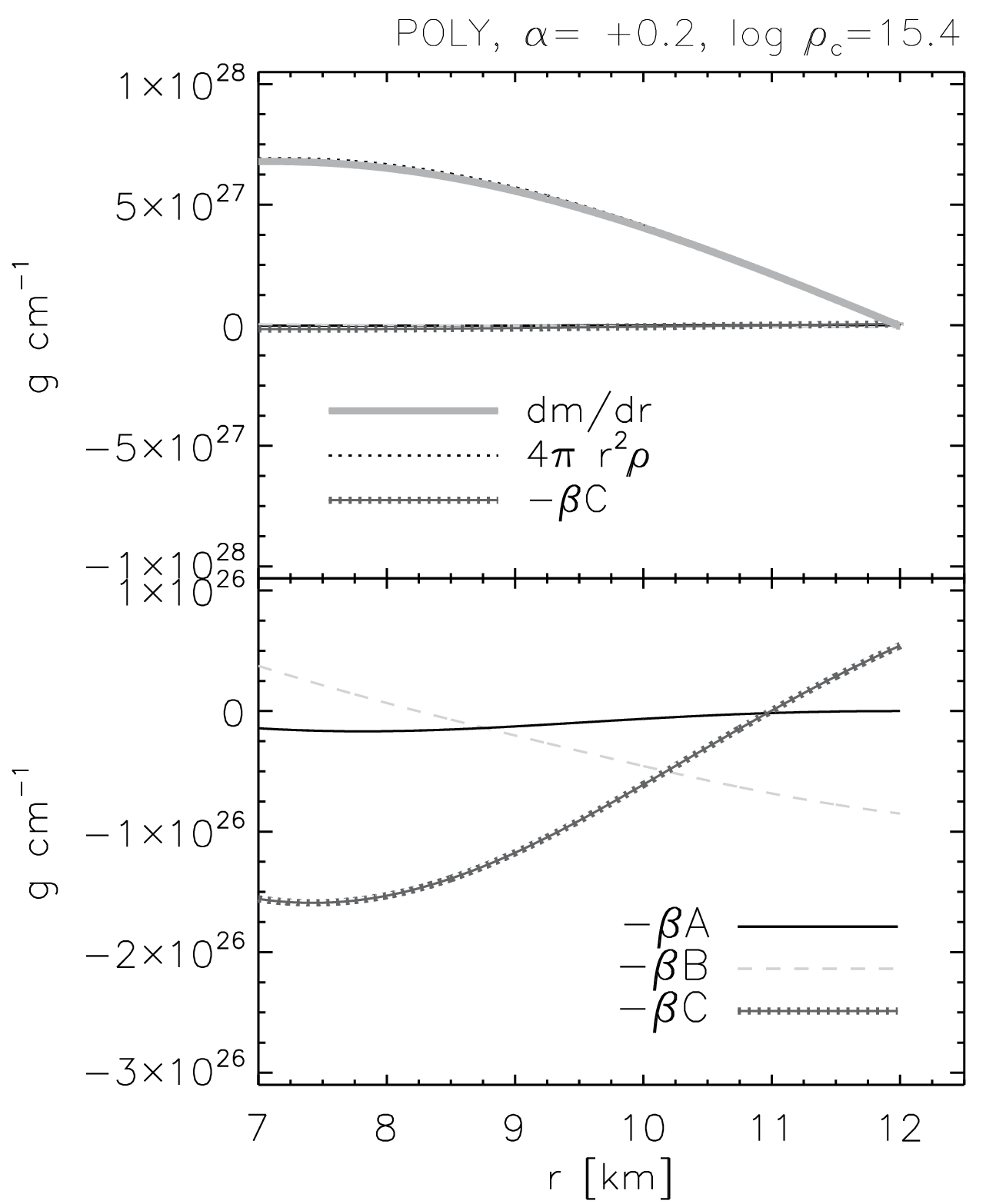}
\caption{Profile of the mass gradient, ${\rm d}m/{\rm d}r$ (thick line), close to the surface of the NSs for SLY (left panel) and POLY (right panel) EoSs. Deviations of the mass profile from the GR case are much more importante for the realistic SLY EoS. Note that the dashed line indicates zeroth-order (GR) term and the continuous with plus signs line indicates the contribution of the $C \propto R_0''/R_0$ term. Lower panels zoomed-in to show in detail the contribution of the minor perturbative terms, indicated in the legend.}
\label{terms_mass}
\end{center}
\hfill\includegraphics[width=.33\textwidth]{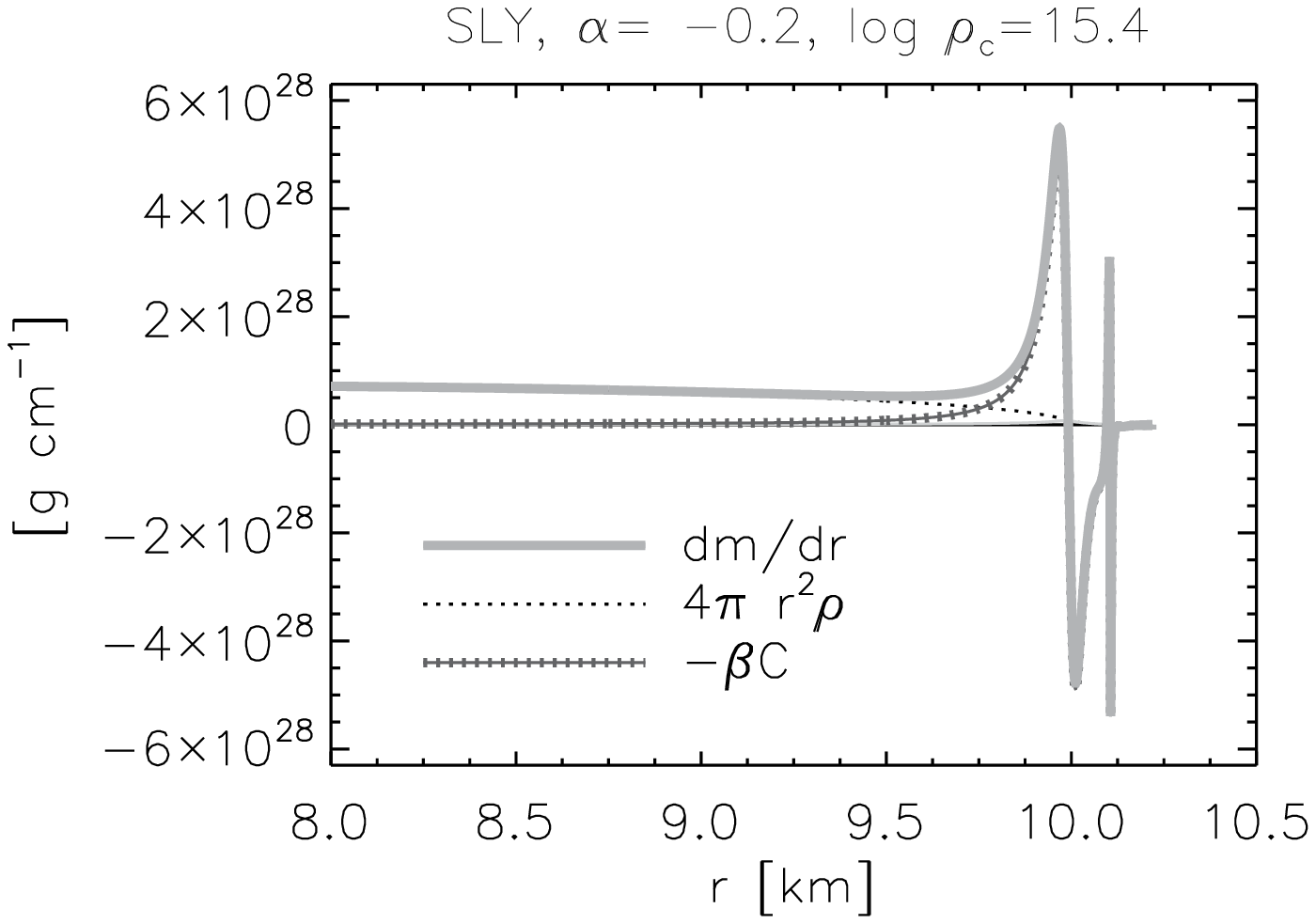}
\hfill\includegraphics[width=.33\textwidth]{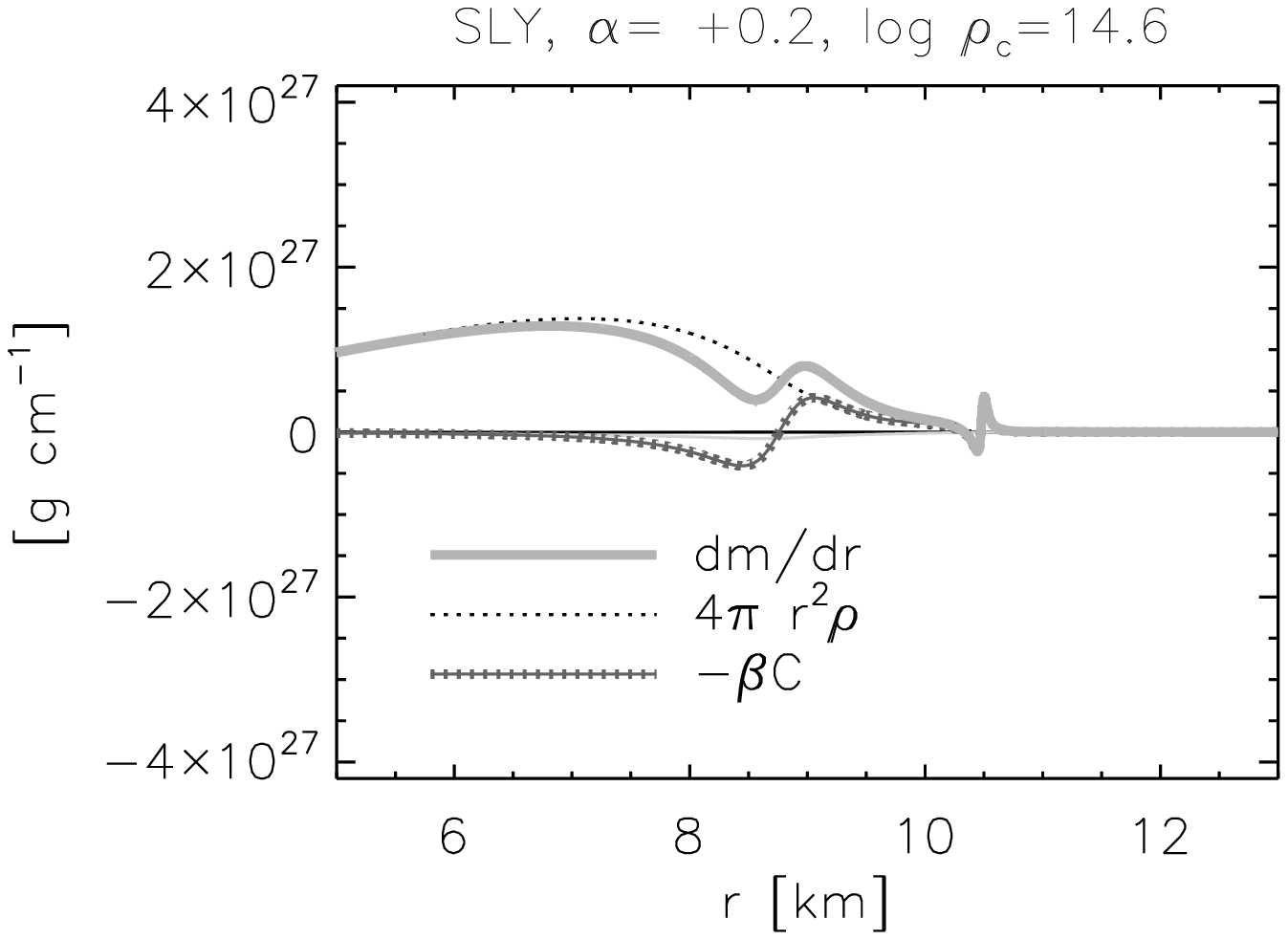}
\hfill\includegraphics[width=.33\textwidth]{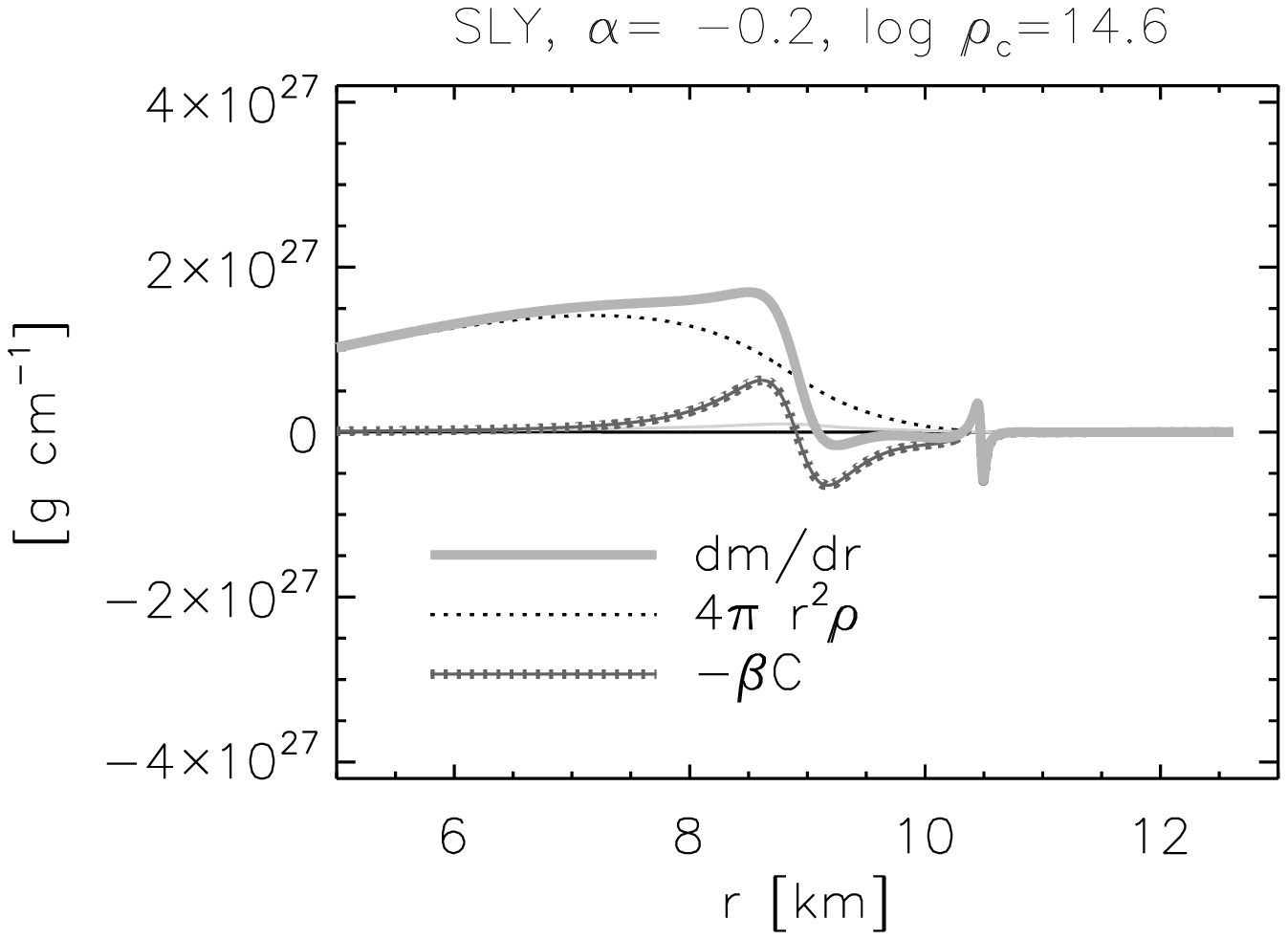}\\
\caption{Same as Figure \ref{terms_mass} for other three SLY cases: $\log \rho_c$~[gr~cm$^{-3}$] = 15.4 (left) and 14.6  (center and right). The central and right plots compare the effects of changing the sign of $\alpha$, which is indicated in ~km$^2$ units. The same values, i.e. $\alpha=-0.2$ (left) and +0.2  (Fig \ref{terms_mass}, left pannel) are shown for a large central density.}
\label{mass_SLY}
\end{figure*}

To further analyse the origin of the deviations from the GR we explore the contribution to ${\rm d}m/{\rm d}r$ of the four terms involved in equation (\ref{Mtov}), which we call: $4\pi r^2 \rho$ (GR) and $A$, $B$, $C$ (perturbative terms).  In Figure \ref{terms_mass} we present each term contribution when $\alpha=+0.2$ ~km$^2$ for SLY (upper-left panel) and POLY (upper-right panel) EoSs, in the case of a high central density star ($\log \rho_c$~[gr~cm$^{-3}$] = 15.4). For the SLY EoS and for values of radii $r\lesssim 9.5$~km, ${\rm d}m/{\rm d}r$ is dominated by the GR term.  Closer to the NS surface, for $r>9.5$~km, the term $C\propto R_0''/R_0$ becomes dominant, causing the fluctuation in the mass profile. On the contrary, for the POLY EoS, which derivatives are strictly bounded smooth functions, this counter-intuitive effect is not present and the trend of ${\rm d}m/{\rm d}r$ is dominated by the GR term, with very small modifications due to the perturbative terms. In the lower panels of Figure \ref{terms_mass} we zoom-in the upper panels to show the behaviour of the minor perturbative terms.

In Figure \ref{mass_SLY} we extend this analysis to compare the behaviour of the mass derivative in four SLY cases: $\log \rho_c$~[gr~cm$^{-3}$] = 15.4 and 14.6 for $\alpha=+0.2$ and --0.2~km$^2$. 
For low central density stars, the effect of the second-order derivative ($C$ term), is relatively less important than in the high central density case.
The fluctuations in this term occur in the density range where the second order logarithmic derivatives of the EoS become relevant, which in this case corresponds to a wider radial band 7--10.5~km, as the density changes in smoother way than in the high central density case. In the latter the density drastically decays in a narrow and superficial range from 9.7 to 10.2~km.

In all cases studied here, the other terms, namely $A$ and $B$, which correspond to $\sim R_0$ and $\sim R_0'$ contributions, respectively, are orders of magnitude lower. 

\section{Conclusions}

With the aim to investigate whether $f(R)$-theories are viable to describe astrophysical scenarios like NSs, we have studied the particular $R$-squared case using both simplified and realistic EoS. We have followed the general steps presented by \citet{Cooney2010} and \citet{Arapoglu2011} using a perturbative approach applied to solve the fourth order field equations. 
Concerning the mass-radius $(M_\star-R_\star)$ relations, we have obtained results consistent with former studies, finding that for the highest absolute values admitted for the $\alpha$ parameter, $f(R)$-theories can accommodate heavier NSs than GR for every EoS. In this sense, it is important to remark that there is no agreement on the maximum mass achievable by NSs before they collapse to black holes, based on the uncertainties present on the behaviour of nuclear matter at the highest densities through their EoS. This problem can not be split out from our lack of understanding of gravity \citep{Wen2011}.

Notwithstanding, our most notorious result concerns the details of the internal structure of NSs considering the largest acceptable value for the $\alpha$ parameter (i.e. the stronger perturbation allowed to GR by the actual constrains). We find that the behaviour of the metric, which in $R$-squared gravity depends not only on the EoS, but also on its higher-order derivatives, leads to a region where the mass enclosed decreases with the radius. 

Despite the fact that, in the frame of GR, this effect could only be explained by means of a negative-density fluid, in $f(R)$ theories, it arises as a natural consequence of the coupled space-time geometry and matter content. Adopting a simple polytropic EoS, with strictly bounded higher-order derivatives, no anomalous behaviours of the internal profiles of the NS structure are observed, and the final effect of modified gravity is reduced.

We emphasize that the uncertainties on the adopted EoS could have an enhanced effect on the solutions. Therefore we add a word of caution, as it remains unclear whether the spikes in the mass profiles arise as a consequence of the analytical representation of the EoS, the perturbative approach or the geometry of \emph{squared}-gravity. In any case, our results raise new questions in the topic. Further research is needed to disentangle these different possibilites.

Finally, we suggest that a thorough study of the stability of the calculated structures would be very important to ensure that these particular configurations can be realized under $R$-squared gravity. Such a work, as well as the study of the astrophysical implications, is beyond the scopes of this paper.


\begin{acknowledgements}
We thank Dr. Arapo{\u g}lu for explanations. The authors appreciate helpful comments from Prof. Santiago E. P\'erez Bergliaffa. M.O. acknowledge support by the Argentine Agency CONICET and ANPCyT through grants PICT 2010-0213/ Prestamo BID and PICT-2007-00848. G.E.R. was supported by PIP 2010-0078 (CONICET) and the Spanish Ministerio de Innovaci\'on y Tecnolog\'ia under grant AYA 2010-21782-C03-01. 
\end{acknowledgements}




\bibliography{NSfR}

\end{document}